\documentclass[conference]{IEEEtran}

\usepackage{cite}
\usepackage[pdftex]{graphicx}
\usepackage{amsmath}
\usepackage{algorithmic}
\usepackage{array}
\usepackage{caption} 
\usepackage{amsmath}
\newtheorem{theorem}{Theorem}

\newtheorem{claim}{Claim}
\newtheorem{definition}{Definition}

\begin{document}
\IEEEoverridecommandlockouts

\title{Using data compression and randomization to build an unconditionally secure short key cipher
}
\author{\IEEEauthorblockN{Boris Ryabko}
\IEEEauthorblockA{Federal Research Center for Information and Computational Technologies
 of SB RAS\\ 
Novosibirsk state university
\\Novosibirsk, Russian Federation \\Email: boris@ryabko.net\\}
}

\maketitle

\begin{abstract}
We consider the problem of constructing an unconditionally secure
 cipher for the case when the key length is less than the length of the encrypted message. (Unconditional security means that a computationally unbounded  adversary cannot obtain  information about the encrypted message  without  the key.) 
In this article, we propose data compression and randomization techniques combined with entropically-secure encryption. The resulting cipher can be used for encryption in such a way that the key length does not depend on the entropy or the length of the encrypted message; instead, it is determined  by the required security level. 
\end{abstract}

\textbf{keywords:} information theory, cryptography, perfect security,  entropic security, entropically-secure symmetric encryption scheme,  indistinguishability,  data compression, randomization, Shannon code.

\section{Introduction}
The concept of unconditional secrecy was presented in the seminal article by C.~Shannon, where he also showed that a one-time pad (or Vernam cipher) is unconditionally secure \cite {sh}. In particular, unconditional secrecy means that a computationally unbounded  adversary cannot obtain any information about an encrypted message  without  a key. It is clear that this property is highly desirable, but if the one-time pad is used to encrypt a message, the length of the key used must be at least the length of the message (or, more precisely, its Shannon entropy). This requirement is too strict for many applications, and there are many approaches   for creating secure ciphers with short or low entropy keys, see 
 \cite{fool,do,ca,med,ry1,ho1,ho2,ry-vernam}. 
One such approach was suggested by C.~Shannon in \cite {sh}, who described  ideal cipher systems where a computationally unbounded  adversary "does
not obtain a unique solution to the cipher but is left with many alternatives, all
of reasonable probability". 
He built a theory of ideal ciphers and described some of them for the case when the probability distribution of encrypted messages is known. Later, ideal systems were proposed for the case of unknown statistics, see \cite{ry1}.

A very interesting and promising new way of building secure systems with short keys is the so-called entropic security, proposed by Russell and Wang  \cite{fool} and developed by Dodis and Smith \cite{do}. This cipher uses the entropy of the original message in such a way that the key length should be roughly the difference between the message length and its min-entropy  (the exact definition will be given below).  Unlike the ideal cipher and some others, this cipher is guaranteed to hide the values of any function of the original message, and not just some of its symbols.
 (Note that the use of the entropy message to improve the strength of the cipher was also used in \cite{ry1,ry-vernam}.) 

The notion of 
 an entropy-secure symmetric encryption scheme is extremely important for cryptography because, as Dodis and Smith \cite{do} said,
   "Russell and Wang showed how one can construct entropically-secure symmetric encryption schemes with keys much shorter than the length of the input, thus circumventing Shannon’s famous lower bound on key length".

Another way to construct a short key cipher is the so-called honey cipher, proposed by Jewels and Ristenpart \cite{ho1} and developed by Joseph, Ristenpart  and Tang  \cite{ho2}.
In some ways, the 
 honey cipher is like the ideal cipher: a computationally unlimited adversary has many possible  highly probable decryptions.  
 Li, Tang and  Zhang  \cite{fo2} developed and combined the ideas of the honey cipher and the entropy ciphers to create a new class of easily implementable short key codes. In a sense, the idea of preprocessing an original message in order to increase its entropy is being developed and widely used in their methods. 

Data compression and randomization are two methods of preprocessing the original message that have been used for centuries in cryptography \cite {sh, gu}.  Moreover, 
 homophonic coding can be used to compress and randomize together \cite{gu, rf}. The goal of both transformations is to make 
  the probability distribution of the original messages closer to the uniform  one (see an overview in  \cite{enc}).  Interestingly, both transformations have been successfully applied to some cryptographic problems: they were used to extract randomness \cite {ne, el, rm} and to build an ideal steganographic system \cite {rr}.

In this article, we combine  entropically-secure encryption with the suggested compression and data randomization techniques. Using compression and randomization,  the original message is  transformed  in such a way that the difference between its length and its min-entropy is constant. 
This make it possible to apply  an entropically-secure  cipher so  that the key length is independent of the entropy or the length of the message   (but depends on the required security level).  

\section{Definitions and preliminaries}

We consider the problem of symmetric encryption, when there are  two parties Alice and Bob and Alice wants  to securely transmit a message $ M  $ to Bob, where $M \in  \{0,1\}^n $, $n\ge 1$,     obeys a certain probability distribution $p$ defined on the set $\{0,1\}^n$. Alice and Bob   have a shared secret key $ K = K_1 ... K_k $, which can be much shorter than the length of $ M $, that is $ k \ll  n $.  Alice encrypts $ M $ with $ K $ and possibly some random bits, and
obtains the word $cipher(M,K)$. Then she  sends it to Bob, who decrypts the received  message and obtains $M$.  In addition, there is a computationally unlimited adversary Eve who does not know $ M $ and $ K $, but  knows the probability distribution $p$ and  wants to find some information about $ M $ based on the encrypted message.   

Russell and Wang  \cite {fool} suggested  a definition of the entropic security which was generalised by Dodis and Smith \cite{do} as follows: 
A probabilistic map $Y$ is said to hide
all functions on $\{0,1\}^n$ 
 with leakage $\epsilon$ if, for every adversary $A$, there exists some adversary $\hat{A}$ (who does not know $ Y(M)$) 
such that for all functions $f$, 
\begin{equation}\label{entrsec} 
| \, Pr\{A( Y(M) ) \, = f(M) \}  \, - Pr\{  \hat{A}(\,)\, = f(M) \} \, | \,\le\, \epsilon .
 \end{equation}
 (note that $\hat{A}$ does not know $ Y(M)$  and, in fact, she guesses the meaning of the function $f(M)$.) 
 In what follows the probabilistic map $Y$ will be $cipher(M,K)$
 and $f$ is a map
 $f: \{0,1\}^n \to \{0,1\}^*$.
\begin{definition}{ }  
The map $Y ()$ is called $(t, \epsilon$)-entropically secure if $Y ()$ hides all functions on $\{0,1\}^n$, whenever the min-entropy of  the probability distribution  $p$ is at least $t$, where min-entropy $h_{min}(p)$ is as follows: 
 \begin{equation}\label{entr}
h_{min}(p) =  - \log \, \, \max_{a \in A} \, p(a) \, .
 \end{equation}
 \end{definition}
(Here and below $\log = \log_2 \,\, $.)

Note, that in a  sense the definition 1  is a generalisation  of the Shannon notation of the perfect security.  Namely, if we take $\epsilon = 0$ and $Y$  $= cipher(M,K)$ and $f(x) = x$, 
we obtain that for any $M$ 
$$ | Pr\{{A}( cipher (M,K) ) = M \}  - Pr \{ \hat{A} (\,\,)  = M \} \,| \, = \, 0 $$
(So, $A$ and $\hat{A}$ obtained the same result, but $A$
estimates  the probability based on $cipher (M,K) $, whereas $\hat{A}$ does it without knowledge of  $cipher (M,K) $).
   So, the entropic security (\ref{entrsec}) can be considered as a generalisation of the Shannon's perfect secrecy. 
  
  The following theorem of Dodis and Smith \cite{do} is a generalisation of the results of Russell and Wang  \cite {fool}.
  \begin {theorem}\label {th1} (\cite{do})  Let there be an alphabet   
  $ \{0,1\}^n, n >0,$ with a probability distribution $p$. Then there exists an efficient   $(h_{min}, \epsilon)$-entropically secure
cipher with the $k$-bit key where 
\begin{equation}\label{maineqv}
k = n - h_{min}(p)+ 2  log (1/\epsilon) +2.
\end{equation}
 \end {theorem}
 
 Another important notion is that of 
  indistinguishability: 
 \begin{definition}\label{indis} ( Dodis and Smith \cite{do}.) 
 A randomized map $Y ()$ is $(t, \epsilon)$-indistinguishable if there is a random variable $ G$
such that for every distribution on messages $\bf{M}$ over $\{0, 1\}^n$ with min-entropy at least $t$, we have
$$SD(Y (M), G)  \le \epsilon,$$
where for two distributions $A,B$  
$$ SD(A,B) = \frac{1}{2} \sum_{M \in \bf{M}}  | Pr\{A = M\} - Pr\{B = M\} | \, .
$$
  \end{definition}

 Dodis and Smith \cite{do} showed that entropic security and indistinguishability are deeply connected:
 \begin{theorem}\label{2}    ( \cite{do} .) 
  Let $Y$ be a randomized map with inputs of length $n$. Then
 
1. $(t, \epsilon)$-entropic security for predicates implies $(t -1, 4\epsilon)$-indistinguishability.

2. $(t -2, \epsilon)$-indistinguishability implies $(t, \epsilon/8)$ -entropic security for all functions when $t \ge 
2 \log (1/\epsilon) + 1$.
\end{theorem}
So,   both notions 
are equal up to small changes in the parameters.

\section{The suggested method}

We can see  from Dodis and Smith Theorem 1 that the choice of the 
length of the key $k$ depends significantly  
on the min-entropy of the probability distribution; specifically,   $ k \ge n - h_{min} + 2 \log (1/\epsilon) + 2$, where $n$ is the length of a ciphered message.  

In this paper we suggest the following two-step cipher: first, transfer (encode) the messages  $M$ in such a way that difference between $n$ and $h_{min}$ is a small constant, and then apply an entropically secure cipher with $k = 2 \log (1/\epsilon) + O(1)$. 
The first transformation will be based on methods of source coding (or data compression)  and randomisation.  Both algorithms  are described in two following short sections.
\subsection{Lossless codes}
\subsubsection{Shannon code and its generalisations}
Let there be an alphabet $A= \{a_1, ... , a_L\}$ with probability distribution $p(a)$ and let 
$p(a_1) \ge p(a_2) \ge ... \ge p(a_L) > 0$. Define $Q_1 = 0, 
Q_t = \sum_{i=0}^{t-1} p(a_i)$, $t=2, ..., L$ and let $\hat{Q_i}$ be a   presentation of 
$Q_i$ in binary system  as an infinite $\{0,1\}$ word with finite number of ones and without the initial  $0$. 
  (That is, 
$1/2 = 100000 ... ,  1/3 = 010101.... $.) 
The codeword  $\hat{\lambda}(a_i)$ for symbol $a_i$  is chosen to be the first $\lceil \log (1/p(a_i) \rceil $ 
  binary digits in 
  $\hat{Q_i}$, $i=1, ... , L$. 
  It is clear that, 
  \begin{equation}\label{sh-c-in}
|\hat{\lambda}(a_i)| = \lceil \log (1/p(a_i) )\rceil  \, .
\end{equation}

  For example, let $A = \{a_1, a_2, a_3\}$ and $p(a_1) = 13/16, p(a_2) = 1/8, p(a_3) = 1/16$.
  Then, $\hat{\lambda}(a_1) = 0,$  $\hat{\lambda}(a_2) = 110, $ $\hat{\lambda}(a_3) = 1111$. Clearly, these codewords can be made shorter as follows:
$\lambda(a_1) = 0,$ $\lambda(a_2) = 10,$ $\lambda(a_3) = 11.$
This procedure for removing extra digits can be described using binary trees.  It is known that the Shannon code can be represented as a binary tree, the branches of which correspond to codewords. In this tree, the left child is marked with 0, and the right child is 1. If some node has one child, it is removed, and the corresponding digit from the corresponding codeword is also removed. The obtained code we denote as $\lambda_{Sh}$ and derive from (\ref{sh-c-in})  the following:
  \begin{equation}\label{sh}
| \lambda_{Sh}(a_i)|  \le \lceil \log (1/p(a_i) ) \rceil  \, \le  \log (1/p(a_i) ) +1.
\end{equation}
Also, it is known that the set of codewords $\lambda(a_1), ... , \lambda(a_L)$ is  prefix-free. (Recall that, by definition, a set of words $U$ is  prefix-free if for any  $u,v \in U$ neither $u$ is a prefix of $v$ nor $v$ is a prefix of $u$.) 

Note that, for any sequence $x_1 x_2  .... x_n, n\ge 1,$ from the alphabet $A$ and a prefix-free code $\lambda$ the encoded sequence   $\lambda(x_1) \lambda(x_2) ...  \lambda(x_n) $ can be decoded to $x_1 x_2  .... x_n$ without errors. Such a code $ \lambda $ is called lossless code. Hence, any prefix-free code is a lossless one.

Note the the ``initial'' code $\hat{\lambda}(a_i)$ has the same properties as a modified $\lambda$, that is, it is the prefix-free and (\ref{sh}) is valid. (That is why we do not describe the transformation of $\hat{\lambda}$   to $\lambda$   in detail and do not estimate its complexity.) 

\subsubsection{Trimmed codes} 
Let   $\lambda$ be a   lossless code for letters from $A$.  
Consider the following probability distribution $p(a_1) =1/2, p(a_2) = 1/4,  ... ,
p(a_{L-1}) = p(a_L) = 2^{-(L-1)}$. From the description of the Shannon code we can see that 
$|\lambda_{Sh}(a_L) )| = {L-1}$.

In the following applications, the complexity of the cipher will largely depend on the lengths of the codewords. Thus, it will be convenient to use codes for which the length of the code of any letter does not exceed $ \lceil \log L \rceil + 1 $ for any probability distribution  (instead of $ L-1 $ as in the previous example).  It is also worth noting that it will be shown later that one extra bit of the length of the codeword can add at most 1 extra bit of the length of the encryption key.

 We call such codes as trimmed and define one of  them as follows:
 if $\lambda$ is a code then  
\begin {equation}\label{tr}
\lambda^{tr}(a_i) =
  \begin{cases}
    0\, \lambda(a_i)       & \quad \text{if } |\lambda(a_i) |\le \lceil \log L \rceil \\
    1\,  bin_{\lceil \log L \rceil}(i) & \quad \text{if }   |\lambda(a_i) | > \lceil \log L \rceil  \, ,
  \end{cases}
\end{equation}
where $ bin_{\lceil \log L \rceil}(i)$ is a binary presentation of $i$ whose length is $\lceil \log L \rceil$. (For example, $bin_3(3) = 011$).   We see that the maximal codeword length is not greater than $ \lceil \log L \rceil + 1 $. Also, note that for any prefix-free code the maximal codeword length is not  less  than $\lceil \log L \rceil $.

Let us explain how to decode. First,  the decoder reads the first binary letter. If it is $0$, the decoder uses the codeword of the code $\lambda$ in order to find the encoded letter. If the first letter is $1$, the next $\lceil \log L \rceil$ letters contain the binary decomposition 
 of $i$, i.e. the letter is $a_i$.

If the trimmed code is built based  on the Shannon code, from (\ref{tr}) and (\ref{sh}) we obtain 
\begin{equation}\label{sh-tr}
| \lambda^{tr}_{Sh}(a_i)|  \le \lceil \log (1/p(a_i) ) \rceil   +1 \, \le  \log (1/p(a_i) ) +2 .
\end{equation}

\subsection{Randomised prefix-free codes}
Let $\lambda$ be a prefix-free code for the alphabet $A$ and $$l = \max_{i=1,...,L} |\lambda(a_i)| \, .$$
The randomized code $\rho_\lambda$  
 maps letters from the alphabet $A$ to the set 
$\{0,1\}^l$ defined as follows.
\begin {equation}\label{ro-la}
\rho_\lambda(a_i) = \lambda(a_i) \,
    r^i_{|\lambda(a_i)|+1} r^i_{|\lambda(a_i)| + 2} ... r^i_l  \, , \end{equation} 
where  $r^i_{|\lambda(a_i)|+1}, r^i_{|\lambda(a_i)| + 2}, ... ,r^i_l $  uniformly distributed and independent random bits (for all $i$).
Let us define a probability distribution $\pi_\lambda$ on   $\{0,1\}^l$ as follows: 
\begin {equation}\label{rand}
\pi_\lambda(y_1y_2 ... y_l) =
   p(a_i) 2^{- (l - |\lambda (a_i) | )} \,   $$ $$     \text{if}  \quad
    y_1 y_2 ... y_{|\lambda(a_i)|} = \lambda(a_i). 
    \end{equation}
    If  for some  $y=y_1 ... y_l $  any $\lambda(a_i)  $ is not a prefix of $y$, then $\pi_\lambda(y) = 0$. 
      \begin {claim}\label {min} 
$h_{min}(\pi_\lambda) = l - \max_{i=1,...,L} (|\lambda(a_i) | - \log (1/p(a_i) )$.
In particular, 
\begin {equation}\label{minSh}
h_{min}(\pi_{\lambda_{Sh}}) > l -1,\,\,
h_{min}(\pi_{\lambda^{tr}_{Sh}}) > l -2 \,.
\end {equation}
\end {claim}  
       Here the first equation follows from the definition of the min-entropy and (\ref{rand}), whereas (\ref{minSh})  follows from (\ref{sh}) and (\ref{sh-tr}).

      
 
\subsection{The cipher}    
Here we describe a cipher with the length of key $2 \log (1/\epsilon ) + O(1) $ which is $(0, \epsilon ) $ entropically secure. It means that this cipher is $\epsilon$ entropically secure for any probability  distribution (i.e. with any min-entropy). This cipher is based on the application of  the entropically secure cipher of Dodis and Smith \cite{do} to the suggested randomized code. We describe the suggested cipher for the $ \lambda_{Sh}^{tr}$ code, but expanding to other prefix-free codes  is straightforward.

The suggested ciphering   is carried out in the following three steps: for a given $\epsilon \in (0,1) $

i) Build a code $\lambda_{Sh}^{tr}$ for the alphabet $ \{0,1\}^n$ with a probability distribution $p$.

ii) Calculate $l$ $= \max_{a \in \{0,1\}^n}$ $|\lambda_{Sh}^{tr}(a)|$  
and probabilities  $\pi_{\lambda^{tr}_{Sh} }(u) $ $, u\in \{0,1\}^l$.
(From (\ref{tr}) we can see that $l \le n+1$.)

iii)  For the alphabet $\{0,1\}^l$   with the distribution $\pi $ build $(l -2, \epsilon) $  entropically secure cipher of Dodis and Smith \cite{do} 
with the $k$-bit key where 
\begin{equation}\label{in}
k =   2  log (1/\epsilon) +4\,  ,
\end{equation}
Note that Dodis and Smith proposed three entropically secure ciphers, any of which can be used; 
 so we do not describe any  particular cipher here.

 From the Dodis and Smith Theorem 1 (see (\ref{maineqv}))   and the estimate of the min-entropy (\ref{minSh})   we can see that such a cipher exists for the distribution $\pi_\lambda$ on $\{0,1\}^l$.  

Strictly speaking, we are dealing with two ciphers. One is applied to the set $\{ \rho_\lambda(M) \}
\subset \{0,1\}^l$,  whereas the second one (the main one) is applied to $\{ M\} \subset 
\{0,1\}^n$, and include data compression and randomization.
 We denote the first one $cipher_{ds}$ and the second  $cipher_{c\&r}$. 
 
 The Dodis and Smith Theorem 1 guarantees the entropic security for the first cipher $cipher_{ds}$, so, we need to prove a similar property for  $cipher_{c\&r}$. 
                The following  theorem describes the properties of this cipher: 
\begin {theorem}\label {my}  
Let any message $M$ belong to $  \{0,1\}^n, n >0,$ and they obey some probability distribution $p$. Let $\epsilon >0$. Then, 
  $cipher_{c\&r}$ is $(0, \epsilon)$  entropically secure with the $k$-bit key with $k =   2  log (1/\epsilon) +4\,$,
that  is, 
for any function $A : \, \{0,1\}^{l} \to \{0,1\}^*$ and $f: \, \{0,1\}^n \to \{0,1\}^*$ 
$$
| Pr \{ A(cipher_{c\&r} (M )  = f(M) \} - Pr \{\hat{A}(\,) = f(M) \} | \le \epsilon ,
$$
where $\hat{A}$  does not use $cipher_{c\&r} (M) $. 

{\it Comment 1}. This means that the  cipher is $\epsilon$- entropically secure
with the secret keys whose lengths do not depend on the probability distribution on the set of the messages $ \{ M\} \subset \{0,1\}^n$.
\end{theorem}
{\it Proof.}   
From Dodis and Smith theorem 1 we see that for any function $g$
$$ 
 | Pr \{ A(cipher_{ds}(v)  = g(v) \} - Pr \{\hat{A}(\,) = g(v) \} | \le \epsilon,
$$
where $v$ is random variable  with distribution  $\{ \pi_\lambda(M), M\in \{0,1\}^n \}$, 
$g$ is any function defined on $\{0,1\}^l$ ($g:  \{0,1\}^l \to \{0,1\}^*)$ and $\hat{A}(\,)$ does not depend on $v$ (to be short, $\lambda = \lambda_{Sh}^{tr}$).
Taking into account that the code $\lambda$ is prefix-free,  we can define such a function $\phi$  that 
for any  $a \in \{0,1\}^n$ and  $u=\rho_\lambda(a) $,    $\,\, \phi(u) = a$.
For any function $f:  \{0,1\}^n\to \{0,1\}^*$  and $M$ consider the function 
$g(\rho_\lambda(M)) = f(\phi (\rho_\lambda(M)) (=f(M))$. The last inequality is valid for this function $g$ and  for $v= \rho_\lambda(M)$, hence 
$$ 
 | Pr \{ A(cipher_{ds}(\rho_\lambda(M) )  =  f(\phi (\rho_\lambda(M)) \} - $$ $$ Pr \{\hat{A}(\,) =  f(\phi (\rho_\lambda(M))  \} | \le \epsilon .
$$
  Taking into account that  $cipher_{c\&r} (M) = cipher_{ds}(\rho_\lambda(M) )$
   and $ f(\phi (\rho_\lambda(M))  = f(M)$, we can see from the latest inequality that  
   $$ 
 | Pr \{ A(cipher_{c\&r}( M))  = f(M) \} - $$ $$ Pr \{\hat{A}(\,) = f(M) \} | \le \epsilon \, .
$$
The theorem is proven. 

We can see from Theorem 2 of Dodis and Smith that
the  cipher  $cipher_{c\&r}$ is 
$2 \epsilon$ indistinguishable.   Below we 
directly prove that $cipher_{c\&r}$ is $\epsilon$ indistinguishable with $k =   2  log (1/\epsilon) +5 \, $. 
\begin {theorem}\label {my2}  Let there be an alphabet    $= \{0,1\}^n, n >0,$ with some  probability distribution. If, instead of (\ref{in}),  the key length $k$ equals $k =   2  log (1/\epsilon) +5 \, $, the cipher is $(0,\epsilon)$ indistinguishable. 
 \end {theorem}
{\it Proof.} 
From Theorem 2 of Dodis and Smith we know that, in fact,  the indistinhuishability is  equal to the entropic security, and it is valid for $cipher_{ds}$, but we are interested in the indistinguishability of the $cipher_{c\&r}$. In order to prove it    suppose that
the Dodis and Smith cipher $cipher_{ds}$  is applied to the words from the set $\rho_\lambda (M)
 \subset \{0,1\}^l$ 
in such a way that it is $(1, \epsilon/4)$  entropically secure, where the length of the key 
equals $2 \log(1/(\epsilon /4))+1 = 2 \log (1/\epsilon) +5$.
From the Dodis and Smith theorem 2 we can see that this cipher is $(2, \epsilon)$ indistinguishable, that is,
$SD(cipher_{ds}  , G) \le \epsilon $, where $G$ is a random variable on $\{0,1\}^l$ which  is independent on $cipher_{ds} $.
  
Define $U_a = \{ cipher_{ds} (\lambda(a)  \, r )\}: r \in \{0,1\}^{l-\lambda(a)} \}$ 
and let $G'(v) $ be defined as follows:
$$Pr\{G' = v\}  = \sum_{w \in U_v} Pr\{ G = w\}.$$
The following chain of equalities and inequalities is based on these definitions and the  triangle inequality for $L_1$:
$$SD(cipher_{c\&r}, G') =  $$ $$\frac{1}{2 }\sum_{u \in \{0,1\}^n } | Pr\{cipher_{c\&r}=u\} - Pr\{ G'=u \}| =
$$
$$  \frac{1}{2 }\sum_{v \in \{0,1\}^n }| \sum_{w \in \{0,1\}^l} 
 Pr\{cipher_{ds}=w\} - Pr\{ G=w \}| \le
$$
$$ \frac{1}{2 }\sum_{v \in \{0,1\}^n } \sum_{w \in \{0,1\}^l} |
Pr\{cipher_{ds}=w\} - Pr\{ G=w \} \, | =
$$
$$
\frac{1}{2 } \sum_{w \in \{0,1\}^l} |
Pr\{cipher_{ds}=w\} - Pr\{ G=w \} \, |  = 
$$
$$
SD(cipher_{ds}, G) \le \epsilon \, .
$$
So, $SD(cipher_{c\&r}, G') \le  \epsilon $.

Theorem is proven.

\section{Conclusion}
As we have seen the suggested cipher is useful for a message distribution with any min-entropy, whereas ciphers of Russell, and Wang as well as  Dodis and Smith are useful if $k = n  - h_{min} +  2 \log (1/\epsilon ) $ is less than $n$, where $n$ is the length of the ciphered message and $k$ is the length of the key (otherwise, one can use the one-time-pad with $k=n$ and, hence, shorter $k$).

Let us consider the choice of the data compression code. We used the trimmed Shannon code, while there are other efficient source codes, among which we mention versions of  Huffman code,  Fano (or Shannon-Fano) codes, Rissanen arithmetic code, and a few others. Taking into account that the difference in  length of the code of the trimmed 
Shannon code and the minimum possible is not more than 2, and the same is true for the length of the secret key of the corresponding ciphers, we consider only the trimmed Shannon code.
On the other hand, it seems that different source codes can be useful in the case of an unknown probability distribution of messages, as well as some other situations.

\section*{Acknowledgment}
Research  was supported  by  Russian Foundation for Basic Research
(grant no. 18-29-03005).


\end{document}